\documentclass[bib]{statapress}

\usepackage{amssymb}
\usepackage{amsmath}
\usepackage{graphicx}
\usepackage{subfigure}
\usepackage{adjustbox}
\usepackage[a4paper,pdftex]{geometry}
\usepackage{mathtools}

\usepackage[crop,newcenter,frame]{pagedims}

\usepackage{sj}

\usepackage{epsfig}

\usepackage{stata}

\usepackage{shadow}

\sjsetissue{$vv$}{$ii$}{$mm$}{$yyyy$}

\begin{document}


\inserttype[st0001]{article}
\author{S. Lee, S. Lee, J. Owusu, and Y. Shin}{%
  Seojeong Lee\\Department of Economics\\Seoul National University\\Seoul, Korea\\s.jay.lee@snu.ac.kr
  \and
  Siha Lee\\Department of Economics\\McMaster University\\Hamilton, Ontario, Canada\\lees223@mcmaster.ca
  \and
  \\
  Julius Owusu\\Department of Economics\\McMaster University\\Hamilton, Ontario, Canada\\owusuj4@mcmaster.ca
  \and \\
  Youngki Shin\\Department of Economics\\McMaster University\\Hamilton, Ontario, Canada\\shiny11@mcmaster.ca
}

\title[csa2sls: Complete subset averaging 2SLS]{csa2sls: A complete subset approach for many instruments using Stata}

\maketitle

\begin{abstract}
We develop a Stata command \texttt{csa2sls} that implements the complete subset averaging two-stage least squares (CSA2SLS) estimator in \citeb{lee2021complete}. The CSA2SLS estimator is an alternative to the two-stage least squares estimator that remedies the bias issue caused by many correlated instruments. We conduct Monte Carlo simulations and confirm that the CSA2SLS estimator reduces both the mean squared error and the estimation bias substantially when instruments are correlated. We illustrate the usage of \texttt{csa2sls} in Stata by an empirical application.

\keywords{\inserttag, \texttt{csa2sls}, many instruments, complete subset averaging, two-stage least squares.}
\end{abstract}

\section{Introduction}

The two-stage least squares (2SLS) estimator is one of the most widely used methods in applied economics. Theoretically, the optimal instrument can be achieved by the conditional mean function of the first-stage regression. However, in practice, practitioners working with a finite sample face a crucial question of how many instruments one should use, especially when there are many instruments available. This is partly due to the well-known trade-off between bias and variance when the number of instruments increases. \citet{donald2001choosing} shows this point clearly by a higher-order Nagar expansion and proposes to choose the optimal number of instruments that minimizes the mean squared errors. \citet{kuersteiner2010constructing} proposes a model averaging approach for the first stage regression and shows that it achieves the optimal weight. These other approaches, however, either require the practitioner to know the order of importance among instruments \citep{donald2001choosing} as the method chooses the first few important instruments, or the practitioner needs to estimate the optimal weights for the instruments \citep{kuersteiner2010constructing}.

As an alternative, \citet{lee2021complete} proposes a model averaging approach that uses all size-$k$ subsets of the set of available instruments in a cross-sectional regression model. This new approach is named as the complete subset averaging two-stage least squares (CSA2SLS) estimator. One advantage of the CSA2SLS estimator is that since it uses all subsets, it does not require knowledge of the order of importance among instruments. Furthermore, averaging models using equal weights reduces potential efficiency loss in finite samples. This is because when estimated weights (instead of equal weights) are used, these become additional parameters in the model and therefore, cause inefficiency when the number of models to be averaged is large. 

We develop a Stata command \texttt{csa2sls} that implements the CSA2SLS estimator. It selects the optimal number of subset size $k$ that minimizes the approximate mean squared errors. Since the size of the complete subset grows at the order of $2^K$, where $K$ is the total number of instruments, CSA2SLS is computationally intensive. To alleviate such a computational burden, the command \texttt{csa2sls} includes options for subsampling and a fast-but-memory-intensive method.

The remainder of the paper is organized as follows. 
In Section 2, we introduce the complete subset averaging two-stage least squares (CSA2SLS) estimator in \citet{lee2021complete}.
Section 3 explains the command \texttt{csa2sls}.
Section 4 shows results from Monte Carlo experiments which numerically illustrates how the CSA2SLS estimator alleviates some of the issues that arise from many instruments.
We provide an empirical application of \texttt{csa2sls} in Section 5. 
Section 6 concludes.

\section{CSA2SLS Estimator}
In this section, we explain the key idea of the CSA2SLS estimator in \citet{lee2021complete}. Heuristically speaking, we estimate the first-stage predicted value by model averaging and apply the 2SLS estimation with those predicted values. Given a total of $K$ instruments, we consider all subsets composed of $k$ instruments. We compute a simple average of predicted values across models and the 2SLS estimator follows immediately. The optimal $k$ is selected by minimizing the approximate mean squared errors criterion, which will be explained in detail below.  

To be concrete, consider the following model generated from an independent and identically distributed (\emph{i.i.d.}) sample:
\begin{align*}
    y_i=\mathbf{Y}_i'\boldsymbol\beta_y + \mathbf{x}'_{1i}\boldsymbol\beta_x + \epsilon_i = \mathbf{X}_{i}'\boldsymbol\beta + \epsilon_i
\end{align*}
 \begin{align*}
     \mathbf{X}_i = \begin{bmatrix}
         \mathbf{Y}_i\\
         \mathbf{x}_{1i}
        \end{bmatrix} = \mathbf{f}(\mathbf{z}_{i}) + \mathbf{u}_i =  \begin{bmatrix}
         E[Y_{i}|\mathbf{z}_{i}]\\
         \mathbf{x}_{1i}
        \end{bmatrix} +  \begin{bmatrix}
         \boldsymbol\eta_i\\
         \mathbf{0}
        \end{bmatrix}, \,\,\, i= 1,...,N 
 \end{align*} where $y_i$ is a scalar outcome variable,  $\mathbf{Y}_i$ is a $d_1\times1$ vector of endogenous variables, $\mathbf{x}_{1i}$ is a $d_2\times1$
vector of included exogenous variables, and $\mathbf{z}_i$ is a vector of exogenous variables (including $\mathbf{x}_{1i}$), $\mathbf{f}(\cdot)$ is an unknown function of $\mathbf{z}$, and $\epsilon_i$ and $\mathbf{u}_i$ are error terms uncorrelated with $\mathbf{z}_i$. Finally, $\boldsymbol\eta_{i}$ denotes an error term when we project the endogenous regressor $\mathbf{Y}_i$ into the space of exogenous variable $\mathbf{z}_i$. Note that $E[\boldsymbol\eta_{i}|\mathbf{z}_{i}] = 0$ by construction.

Let $y = (y_{1},.., y_{N})'$, $\epsilon = (\epsilon_1,...,\epsilon_N)'$, $\mathbf{X} = (\mathbf{X}_{1},..., \mathbf{X}_{N})'$, $\mathbf{f} = (\mathbf{f}_{1},...,\mathbf{f}_{N})'$, and $\mathbf{U} = (\mathbf{u}_{1},...,\mathbf{u}_{N})'$ where $\mathbf{f}_i = \mathbf{f}(\mathbf{z}_{i})$.
The set of instruments has the form $\mathbf{Z}_{K,i} \equiv ( \psi_{1}(\mathbf{z}_{i}),...,\psi_{K}(\mathbf{z}_{i}), \mathbf{x}_{1i})'$, where $\psi_{k}$'s are functions of $\mathbf{z}_{i}$ such that $\mathbf{Z}_{K,i}$ is the collection of $(K + d_2)$ instruments. Note that the total number of instruments $K$ can increase as $N \rightarrow \infty$. We suppress the dependency of $K$ on $N$ for notation simplicity. Let $\mathbf{Z}_K = (\mathbf{Z}_{K,1},..., \mathbf{Z}_{K,N})'$ be the collection of $\mathbf{Z}_{K,i}$. 

Let $M$ be the number of subsets (or models) with $k$ instruments:
\begin{align*}
    M = \dbinom{K}{k} =\frac{K!}{k!(K-k)!}.
\end{align*}
We also suppress the dependency of $M$ on $K$ and $k$. Let $m\in\{1,\ldots,M\}$ be an index of each model and $\mathbf{z}_{m,i}^{k}$ be a vector of instruments in model $m$. Then, the first-stage regression of model $m$ can be written as
 \begin{align*}
     \mathbf{X} = \boldsymbol\Pi_{m}^{k'}\mathbf{Z}_{m}^{k} + \mathbf{u}_{m}^{k}.
 \end{align*} 
 The average predicted value of $X$ is 
 \begin{align*}
    \hat{\mathbf{X}}=\frac{1}{M}\sum_{m=1}^{M}\mathbf{Z}_{m}^{k}\hat{\boldsymbol\Pi}_m^{k},
\end{align*} 
where  $\hat{\Pi}_m^{k}$ is the OLS estimator of ${\Pi}_m^{k}$. Then, the CSA2SLS estimator is defined as
\begin{align*}
    \hat{\boldsymbol\beta} = \big(\hat{\mathbf{X}}'\mathbf{X}\big)^{-1}\hat{\mathbf{X}}'\mathbf{y}.
\end{align*}
Using the projection matrices, we can also write the CSA2SLS estimator as a one-step procedure:
\begin{align*}
    \hat{\boldsymbol\beta} = \big(\mathbf{X}'\mathbf{P}^{k}\mathbf{X}\big)^{-1}\mathbf{X}'\mathbf{P}^{k}\mathbf{y}
\end{align*}
where $\mathbf{P}^k = {M}^{-1}\sum_{m=1}^{M} \mathbf{P}_{m}^{k}$ with $\mathbf{P}_{m}^{k} = \mathbf{Z}_{m}^k\left(\mathbf{Z}_{m}^{k'}\mathbf{Z}_{m}^k\right)^{-1}\mathbf{Z}_{m}^{k'}$.

The optimal subset size $k$ is chosen by minimizing the approximate mean squared error. 
Let $\tilde{\boldsymbol\beta}$ be a preliminary estimator  and $\tilde{\boldsymbol\epsilon} = \mathbf{y}-\mathbf{X}\tilde{\boldsymbol\beta}$. The fitted value of $\mathbf{f}$ is given as 
\begin{align*}
    \tilde{\mathbf{f}} =\tilde{\mathbf{Z}}^k\big(\tilde{\mathbf{Z}}^{k'}\tilde{\mathbf{Z}}^{k}\big)^{-1}\tilde{\mathbf{Z}}^{k'}\mathbf{X}
\end{align*} 
where $\tilde{\mathbf{Z}}^k$ consists of exogenous variables plus the preliminary selection of instruments as described above. Let $\tilde{\mathbf{P}}_Z = \tilde{\mathbf{Z}}^k\big(\tilde{\mathbf{Z}}^{k'}\tilde{\mathbf{Z}}^{k}\big)^{-1}\tilde{\mathbf{Z}}^{k'}$.
 The residual matrix is denoted by $\tilde{\mathbf{u}} = \mathbf{X}-\tilde{\mathbf{f}}$. Define $\tilde{\mathbf{H}} = \tilde{\mathbf{f}}'\tilde{\mathbf{f}}/N$, $\tilde{\sigma}_{\epsilon}^2 = \tilde{\epsilon}'\tilde{\epsilon}/N$, $\tilde{\sigma}_{u\epsilon} = \tilde{u}'\tilde{\epsilon}/N$, $\tilde{\sigma}_{\lambda\epsilon} =  \tilde{\boldsymbol\lambda}'\tilde{\mathbf{H}}^{-1}\tilde{\sigma}_{u\epsilon} $ and $\tilde{\Sigma}_{u} = \tilde{\mathbf{u}}'\tilde{\mathbf{u}}/N$.
Then, the sample counterpart of the approximate mean squared error is given by
\begin{align*}
    \hat S_\lambda(k) = \tilde{\sigma}_{\lambda\epsilon}^2 \frac{k^2}{N} + \tilde{\sigma}_{\epsilon}^2 \big[\tilde{\boldsymbol\lambda}'\tilde{\mathbf{H}}^{-1}\tilde{e}_{f}^{k}\tilde{\mathbf{H}}^{-1}\tilde{\boldsymbol\lambda} - \tilde{\boldsymbol\lambda}'\tilde{\mathbf{H}}^{-1}\tilde{\boldsymbol\xi}_{f}^{k}\tilde{\mathbf{H}}^{-1}\tilde{\boldsymbol\xi}_{f}^{k}\tilde{\mathbf{H}}^{-1}\tilde{\boldsymbol\lambda}\big],
\end{align*} where 
\begin{equation*}
    \tilde{e}_{f}^{k}=\frac{\mathbf{X}'(\mathbf{I} - \mathbf{P}_{k})^{2}\mathbf{X}}{N} + \tilde{\boldsymbol\Sigma}_{u}\big(\frac{2k-tr((\mathbf{P}^{k})^{2})}{N}\big)
\end{equation*}
\begin{equation*}
    \tilde{\boldsymbol\xi}_{\mathbf{f}}^{k}=\frac{\mathbf{X}'(\mathbf{I} - \mathbf{P}_{k})^{2}\mathbf{X}}{N} + \tilde{\boldsymbol\Sigma}_{u}\frac{k}{N} - \tilde{\boldsymbol\Sigma}_{u}
\end{equation*}
\begin{equation*}
    \tilde{\sigma}_{\lambda\epsilon}^2 = (\tilde{\boldsymbol\lambda}'\tilde{\mathbf{H}}^{-1}\tilde{\sigma}_{\lambda\epsilon})^2
\end{equation*}
The preliminary estimator $\tilde{\boldsymbol\beta}$ can be estimated either by using the two-step Malllows's criterion or by adopting the one-step method. See \citet{lee2021complete} for detail. 

\section{The command \texttt{csa2sls}}
\subsection{Syntax}
The generic syntax for the command is as follows:
\begin{stsyntax}
    csa2sls
    \textit{depvar}
    \textit{exogvar (endovar = instvar)}
  [ , \texttt{noprint} \texttt{noconstant} \texttt{r}(\textit{integer}) \texttt{large} \texttt{onestep} ]
\end{stsyntax}
The  four arguments of the command are \textit{depvar} (the dependent variable), \textit{exogvar} (the list of included   exogenous variables), \textit{endovar} (the endogeneous variable) and \textit{instvar} (the list of excluded exogenous variables, i.e.\ instrumental variables). 
\begin{description}
\item \texttt{noprint} specifies estimation without printing the results on screen. The default is to print estimation results in the result pane.
\item \texttt{noconstant}  specifies estimation without an intercept term in the second stage regression. The default is to include a constant term.
\item \texttt{r}(\textit{integer}) specifies a positive integer for the  maximum number of randomly selected subsets when the number of subsets are bigger than \textit{integer}. The default is 100.
\item \texttt{large} turns on the large sample estimation program. When the sample size is large, the average projection matrices may require a large memory size. The \texttt{large} option must be turned on to avoid an insufficient memory issue. The default is not using this option.
\item \texttt{onestep} specifies the preliminary estimation method using the one step approach described in \citeb{lee2021complete}. The default is the preliminary two-step Mallows criterion.
\end{description}
\subsection{Stored results}
 \texttt{ereturn list} command has the following stored information:
\begin{description}
\item Scalars: 
\begin{description}
\item $e(rss)$ equals the residual sum of squares.
\item $e(optimal\_k)$ equals the optimal subset size of instruments.
\item $e(df\_m)$  equals the model degrees of freedom.
\item $e(rmse)$  equals the root mean squared error.
\item $e(mss)$  equals the model sum of squares.
 \item $e(r2)$   equals the R-squared.
 \item $e(r2\_a)$  equals the adjusted R-squared.
 \item $e(chi2)$ equals the  chi-squared.
\item $e(N)$ equals the number of observations.
\item $e(rank)$ equals the rank of $e(V).$
\end{description}

\item Macros: 
\begin{description}
\item $e(Premethod)$ describes the preliminary estimation method. 
\item $e(cmd)$ is the  name of the command. 
\item $e(depvar)$ is the name of the dependent variable.
\item $e(exogr)$ is the name of the exogenous variables.
\item $e(insts)$ is the name of the instruments.
\item $e(instd)$ is the name of the instrumented variables.
\item $e(cmdline)$ is the command line typed by the user.
\item $e(properties)$ specifies the name of the coefficient and variance-covariance matrix. 
\item  $e(predict)$  is the program used to implement \textbf{predict}
\item  $e(constant)$ is \textbf{noconstant} or \textbf{hasconstant} if specified.
\item  $e(clustvar)$  is the  name of cluster variable.
\item  $e(footnote)$  program used to implement footnote display. 
\item  $e(title)$ is the title in estimation output.
\end{description}

\item Matrices: 
\begin{description}
\item $e(b)$ equals the coefficient matrix
\item $e(V)$ equals the variance-covariance matrix

\end{description}
\end{description}

\section{Monte Carlo experiments}
In this section, we conduct Monte Carlo simulation studies focusing on the effect of correlated instruments. An $i.i.d.$ sample $(y_i,Y_i, \mathbf{z}_i)$ is generated from the following simulation design:
\begin{align*}
y_i&=\beta_0+ \beta_1{Y}_i +\epsilon_i\\
{Y}_i&=\boldsymbol\pi'\mathbf{z}_i+ {u}_i,
\end{align*}
where $Y_i$ is a scalar endogenous regressor, $(\beta_0,  \beta_1)$ is set to be $(0, 0.1)$, and $\mathbf{z}_i$ is a $K$-dimensional vector of instruments generated from a multivariate normal distribution $N(0,\boldsymbol\Sigma_z)$. The diagonal elements of $\boldsymbol\Sigma_z$ are set to be 1 and the off-diagonal elements are $\rho_z$. We set each element of $\boldsymbol\pi$ to be $\sqrt{0.1/(K + K(K-1)\rho_z (1-0.1))}$, where 0.1 is the R-squared in the first stage regression. The vector of error terms $(\epsilon_i, {u}_i)$ follows a bivariate normal distribution whose means are zeros and variances are ones. The covariance between $\epsilon_i$ and $\mathbf{u}_i$ is set to be 0.9. In these simulation studies, $K$ varies in $\{5, 10, 15, 20\}$ and $\rho_z$ varies in $\{0, 0.5, 0.9\}$. The sample size is set to be $n=100$ and the results are from 1,000 replications. 

\begin{figure}[htpb]
    \centering
    \caption{Mean Bias and Mean Squared Error}\label{fig:example}
    \subfigure[\centering $\rho_z=0.0$]{{\includegraphics[scale=0.75]{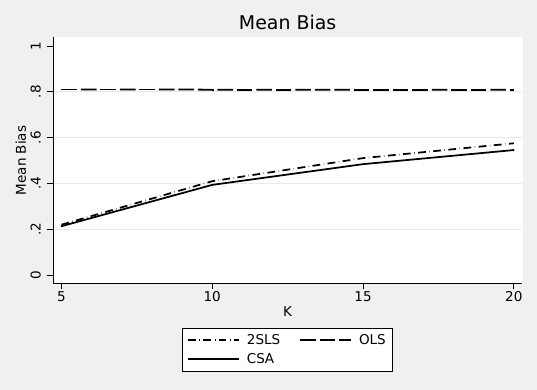} }~{\includegraphics[scale=0.75]{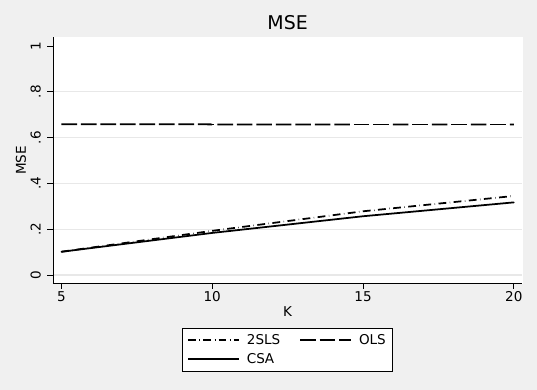} }}
    \\
    \subfigure[\centering $\rho_z=0.5$]{{\includegraphics[scale=0.75]{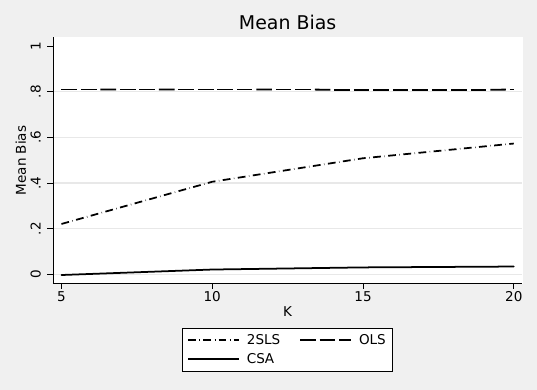} }~{\includegraphics[scale=0.75]{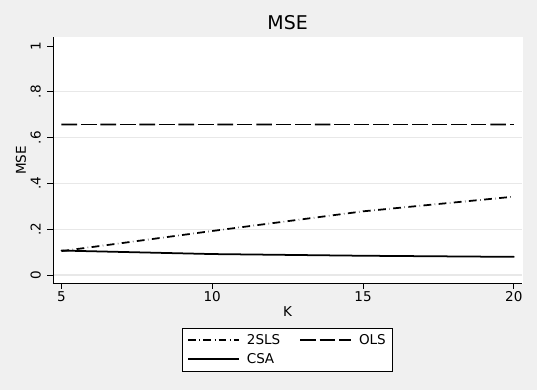} }}
    \\
    \subfigure[\centering $\rho_z=0.9$]{{\includegraphics[scale=0.75]{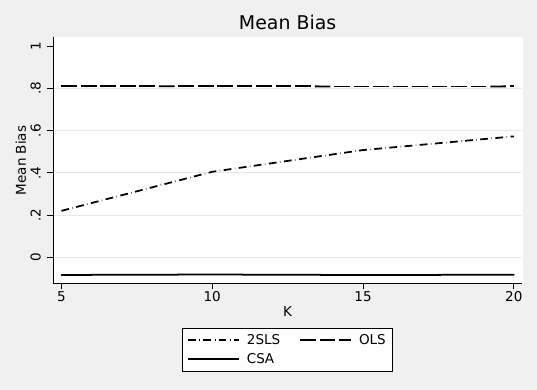} }~{\includegraphics[scale=0.75]{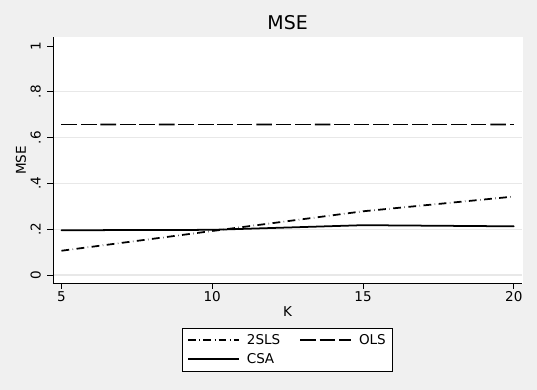} }}
    
\end{figure}

Figure \ref{fig:example} summarizes the simulation results. We report the mean bias and mean squared error (MSE) of CSA2SLS along with the performance of the ordinary least squares estimator (OLS) and the two-stage least squares estimator (2SLS). 
First, the CSA2SLS estimator reduces the bias substantially when instruments are correlated ($\rho_z=0.5, 0.9$). As predicted by theory, the bias of 2SLS increases as $K$ increases. Note that when instruments are independent ($\rho_z=0.0$), the difference in the bias between the CSA2SLS estimator and the 2SLS estimator is small. \citet{lee2021complete} proves that the performance of CSA2SLS will be asymptotically equivalent to that of 2SLS when $\rho_z=0$.

Second, the efficiency loss of CSA2SLS is modest. When instruments are correlated, CSA2SLS achieves lower mean squared errors when $K \ge 10$. Like the bias, the MSE gap between CSA2SLS and 2SLS increases as $K$ increases. It is also worthwhile to note that the MSE of CSA2SLS does not change much over different values of  $K$.
Finally, the OLS estimator performs the worst in these simulation designs.

To sum, the CSA2SLS estimator shows a good finite sample performance as predicted by theory. We also observe the increased bias of 2SLS when there are many instruments. We recommend practitioners to use the CSA2SLS estimator when they have many \emph{correlated} instruments.

\section{Empirical illustration}
In this section we illustrate the usage of \texttt{csa2sls} by an empirical application. In this example, we revisit \citet*{berry1995automobile} and estimate a logistic demand function for automobiles based on pooled cross-sectional data over different markets.

The model is specified as
\begin{align*}
log(S_{i}) -log(S_{0}) & = \alpha_0 P_{i} + \mathbf{X}'_{i}\boldsymbol\beta_0 + \epsilon_{i} \\
P_{i} & = \mathbf{Z}'_{i}\boldsymbol\delta_0 + \mathbf{X}'_{i}\boldsymbol\rho_0 + {u}_{i},
\end{align*}
where $S_{i}$ is the market share of product $i$ with product 0 denoting the outside option, $P_{i}$ is the endogenous price variable, $\mathbf{X}_{i}$ is a vector of included exogenous variables and $\mathbf{Z}_{i}$ is a set of 10 instruments. The parameter of interest is $\alpha_0$ from which we can calculate the price elasticity of demand. 
Note that the optimal subset size $k$ is 9 in this empirical example.

\begin{stlog}
. set seed 2022
{\smallskip}
. insheet using "BLP.csv", comma clear
(54 vars, 2,217 obs)
{\smallskip}
. csa2sls y hpwt air mpd space (price = sumother1 sumotherhpwt sumotherair ///
> sumothermpd sumotherspace sumrival1 sumrivalhpwt sumrivalair sumrivalmpd ///  
> sumrivalspace)
{\smallskip}
Complete Subset Model Averaging 2SLS Regression   Number of obs   =      2,217
                                                  Wald chi2(5)    =     820.64
                                                  Prob > chi2     =     0.0000
                                                  R-squared       =     0.3373
                                                  Root MSE        =     1.1245
{\smallskip}
\HLI{13}{\TOPT}\HLI{64}
           y {\VBAR}      Coef.   Std. Err.      z    P>|z|     [95\% Conf. Interval]
\HLI{13}{\PLUS}\HLI{64}
       price {\VBAR}   -.142563   .0117095   -12.18   0.000    -.1655131   -.1196128
        hpwt {\VBAR}   1.422452    .414676     3.43   0.001     .6097024    2.235202
         air {\VBAR}   .5620958   .1379201     4.08   0.000     .2917772    .8324143
         mpd {\VBAR}   .1579617   .0471821     3.35   0.001     .0654864    .2504369
       space {\VBAR}   2.284253   .1289588    17.71   0.000     2.031499    2.537008
       _cons {\VBAR}  -2.342198   .2673599    -8.76   0.000    -2.866214   -1.818182
\HLI{13}{\BOTT}\HLI{64}
Instrumented:  price
Instruments:   hpwt air mpd space sumother1 sumotherhpwt sumotherair
               sumothermpd sumotherspace sumrival1 sumrivalhpwt sumrivalair
               sumrivalmpd sumrivalspace
{\smallskip}
    . correlate sumother1 sumotherhpwt sumotherair sumothermpd sumotherspace ///
>  sumrival1 sumrivalhpwt sumrivalair sumrivalmpd sumrivalspace
(obs=2,217)
{\smallskip}
{\scriptsize
             {\VBAR} sumoth{\tytilde}1 sumoth{\tytilde}t sumoth{\tytilde}r sumoth{\tytilde}d sumoth{\tytilde}e sumriv{\tytilde}1 sumriv{\tytilde}t sumriv{\tytilde}r sumriv{\tytilde}d sumriv{\tytilde}e
\HLI{11}{\PLUS}\HLI{80}
   sumother1 {\VBAR}   1.0000
sumotherhpwt {\VBAR}   0.9791   1.0000
 sumotherair {\VBAR}   0.6948   0.7039   1.0000
 sumothermpd {\VBAR}   0.9309   0.9341   0.7914   1.0000
sumothersp{\tytilde}e {\VBAR}   0.9902   0.9747   0.6335   0.8862   1.0000
   sumrival1 {\VBAR}  -0.3873  -0.3552   0.0832  -0.1527  -0.4667   1.0000
sumrivalhpwt {\VBAR}  -0.2744  -0.2163   0.1680  -0.0271  -0.3487   0.9532   1.0000
 sumrivalair {\VBAR}  -0.0227   0.0089   0.3275   0.2013  -0.1035   0.8830   0.9168   1.0000
 sumrivalmpd {\VBAR}  -0.1400  -0.0923   0.2531   0.1132  -0.2131   0.9053   0.9456   0.9281   1.0000
sumrivalsp{\tytilde}e {\VBAR}  -0.5178  -0.4797  -0.0277  -0.2790  -0.5909   0.9823   0.9356   0.8144   0.8576   1.0000
}
{\smallskip}
{\smallskip}

\end{stlog}

We also report correlation coefficients among the instruments. We can confirm that the instruments are divided into two groups and that each group's instruments are highly correlated with each other. The correlation coefficient varies from 0.69 to 0.99. 

\section{Conclusion}
In this article, we present the complete subset averaging two-stage least squares estimator and develop the corresponding Stata command \texttt{csa2sls}. The usage of \texttt{csa2sls} is illustrated by an empirical application. The Monte Carlo experiments show that 2SLS is biased when there are many instruments and that CSA2SLS outperforms 2SLS when instruments are correlated with each other. Since CSA2SLS is computationally intensive, an interesting future research question would be to develop a more efficient computation algorithm. An approach based on the stochastic gradient descent (see, for example, \citet{lee2022fast}) can be a possible solution.

\section{Acknowledgement}
We would like to thank the editor and an anonymous reviewer for their valuable comments on this article and for their helpful feedback on the program code.  Shin is grateful for partial support by the Social Sciences and Humanities Research Council of Canada (SSHRC-435-2021-0244).

\bibliographystyle{sj}
\bibliography{sj}

\ifnum 5=1 \def\bibname{Reference}
\else \def\bibname{References} \fi
\begin{thebibliography}{5}
\expandafter\ifx\csname natexlab\endcsname\relax\def\natexlab#1{#1}\fi
\expandafter\ifx\csname url\endcsname\relax
  \def\url#1{\texttt{#1}}\fi
\expandafter\ifx\csname urlprefix\endcsname\relax\def\urlprefix{URL }\fi

\bibitem[{Berry et~al.(1995)Berry, Levinsohn, and Pakes}]{berry1995automobile}
Berry, S., J.~Levinsohn, and A.~Pakes. 1995.
\newblock Automobile prices in market equilibrium.
\newblock \emph{Econometrica} 63(4): 841--890.

\bibitem[{Donald and Newey(2001)}]{donald2001choosing}
Donald, S.~G., and W.~K. Newey. 2001.
\newblock Choosing the number of instruments.
\newblock \emph{Econometrica} 69(5): 1161--1191.

\bibitem[{Kuersteiner and Okui(2010)}]{kuersteiner2010constructing}
Kuersteiner, G., and R.~Okui. 2010.
\newblock Constructing optimal instruments by first-stage prediction averaging.
\newblock \emph{Econometrica} 78(2): 697--718.

\bibitem[{Lee et~al.(2022)Lee, Liao, Seo, and Shin}]{lee2022fast}
Lee, S., Y.~Liao, M.~H. Seo, and Y.~Shin. 2022.
\newblock Fast and robust online inference with stochastic gradient descent via
  random scaling.
\newblock In \emph{Proceedings of the AAAI Conference on Artificial
  Intelligence}, vol.~36,  7381--7389.

\bibitem[{Lee and Shin(2021)}]{lee2021complete}
Lee, S., and Y.~Shin. 2021.
\newblock Complete subset averaging with many instruments.
\newblock \emph{The Econometrics Journal} 24(2): 290--314.

\end{thebibliography}

\begin{aboutauthors}
Seojeong Lee is an associate professor of Economics at Seoul National University

Siha Lee is an assistant professor of Economics at McMaster University

Julius Owusu is a doctoral candidate in Economics at McMaster University.

Youngki Shin is a professor of Economics at McMaster University
\end{aboutauthors}

\clearpage
\end{document}